# Quasi-2D anomalous Hall Mott insulator of topologically engineered $J_{\text{eff}}$ =1/2 electrons

Junyi Yang[1], Hidemaro Suwa[2], Derek Meyers[3,4], Han Zhang[1], Lukas Horak[5], Zhaosheng Wang[6], Gilberto Fabbris[7], Yongseong Choi[7], Jenia Karapetrova[7], Jong-Woo Kim[7], Daniel Haskel[7], Philip J. Ryan[7,8], M. P. M. Dean[3], Lin Hao[6,*], and Jian Liu[1,*]

[1]Department of Physics and Astronomy, University of Tennessee, Knoxville, Tennessee 37996, USA

[2]Department of Physics, University of Tokyo, Tokyo, Japan

[3]Department of Condensed Matter Physics and Materials Science, Brookhaven National Laboratory, Upton, New York 11973, USA

[4]Department of Physics, Oklahoma State University, Stillwater, OK 74078

[5]Department of Condensed Matter Physics, Charles University, Ke Karlovu 5, 12116 Prague, Czech Republic

[6]Anhui Key Laboratory of Condensed Matter Physics at Extreme Conditions, High Magnetic Field Laboratory, HFIPS, Chinese Academy of Sciences, Hefei, Anhui 230031, China

[7]Advanced Photon Source, Argonne National Laboratory, Argonne, Illinois 60439, USA

[8]School of Physical Sciences, Dublin City University, Dublin 9, Ireland

* To whom correspondence should be addressed. E-mail: haolin@hmfl.ac.cn, jianliu@utk.edu

**Abstract**

We investigate an experimental toy-model system of a pseudospin-half square-lattice Hubbard Hamiltonian in [(SrIrO$_3$)$_1$/(CaTiO$_3$)$_1$] to include both nontrivial complex hopping and moderate electronic correlation. While the former induces electronic Berry phases as anticipated from the weak-coupling limit, the later stabilizes an antiferromagnetic (AFM) Mott insulator ground state in analogous to the strong-coupling limit. Their combined results in the real system are found to be an anomalous Hall effect with a non-monotonic temperature dependence due to the self-competition of the electron-hole pairing in the Mott state, and an exceptionally large Ising anisotropy that is captured as a giant magnon gap beyond the superexchange approach. The unusual phenomena highlight the rich interplay of electronic topology and electronic correlation in the intermediate-coupling regime that is largely unexplored and challenging in theoretical modelling.

# Introduction

The concept of topology has revolutionized the understanding of quantum materials. Successful developments have been made within the last decade in describing and discovering a variety of novel topological phases, symmetry-protected states, and Hall effects in non-interacting electronic systems [1-6]. Electronic correlation, on the other hand, remains a profound and challenging problem, where rich and fascinating emergent phenomena have been known or predicted for a long time and yet to be fully understood, such as Mott transition, unconventional superconductivity, and quantum magnetism [7-13]. Fundamental interests arise in systems where both electronic correlation and electronic topology play significant roles [14-20]. Their interplay in this extensive and largely unexplored regime could provide not only new routes to unresolved problems but also opportunities for stabilizing novel quantum states [21, 22].

A key quantity in electronic topology is the Berry phase, which can be expressed as the gauge-invariant phase change of the wave function acquired when a particle circles around a closed loop [4]. This phase change can be conceptualized as a fictitious magnetic field applied to a moving charge carrier, giving rise to an anomalous velocity in transport, such as that causing the anomalous Hall effect (AHE) [5]. In single-particle band structures, the Berry phase often concentrates near band crossing, such as Dirac points, in the momentum space. Many microscopic models encode these effects using complex hopping parameters, with famous examples being the Haldane model [23] and the Kane-Mele model [24, 25] for two-dimensional (2D) topological band insulators on graphene-like honeycomb lattices. In Mott materials, however, electrons are localized in real space and charge hopping is considered as perturbation that gives rise to superexchange interactions [26]. The incompatibility between these two opposite views of the electronic structure highlights the challenge in understanding topological correlated systems [12, 14]. Although the magnetic orders of correlated electrons have been exploited as time-reversal-symmetry-breaking fields for lifting topological bands and Dirac points [27], this mean-field approach remains an effective single-particle method and could fail to predict or account for many-body effects.

An alternative strategy is to experimentally realize toy-model materials where the topology-correlation interplay can be captured and controlled by engineering the complex hopping as well as the correlation strength. The design of such systems thus necessarily requires strong spin-orbit coupling (SOC) [4], which is the source of complex hopping. A proper symmetry configuration is also essential since strong SOC itself does not guarantee nontrivial topology. Meanwhile, the correlation strength should be in the intermediate coupling regime which connects the opposite limits and provides the reconciliation of the two incompatible views. This regime is also the more interesting and yet challenging regime for theoretical descriptions due to the proximity to the Mott transition [7, 28-32] and spatial charge-spin fluctuations associated with excitations into the electron-hole continuum [7, 14, 32, 33]. To fulfill these requirements, we designed and realized an artificial layered iridate [$(SrIrO_3)_1/(CaTiO_3)_1$] to simulate a single-orbital square-lattice Hubbard Hamiltonian in the intermediate coupling regime with SU(2)-symmetry-breaking complex hopping that is implemented by symmetry engineering and in virtue of the strong SOC of the 5$d$ electrons.

## Results

**Pseudospin-half square-lattice Hubbard model**

The single-orbital square-lattice Hubbard Hamiltonian is an iconic model of correlated physics with the $S = 1/2$ $CuO_2$ plane in the cuprates being one of the most profound examples [8, 34]. While SOC is small for Cu, it has been recently recognized that a partially analogous system can be achieved with the $J_{eff}$ = 1/2 state of $Ir^{4+}$ ion confined in a monolayer of corner-sharing $IrO_6$ octahedra in layered iridates [35-40]. A key difference is that, due to the spin-orbit-entangled $J_{eff}$ = 1/2 wave function [37, 41], significant spin-dependent hopping arises when the octahedra rotate and break the local inversion-symmetry of individual Ir-O-Ir bond [42-45]. This is captured by a frozen SU(2) gauge field attached to the hopping parameter of the Hubbard Hamiltonian [42-45], which can be written as

$$H = -t \sum_{<ij>} \sum_{\alpha,\beta} \left[ c_{i\alpha}^\dagger (e^{i\theta d_{ij} \cdot \sigma})_{\alpha\beta} c_{j,\beta} + h.c. \right] + U \sum_i n_{i\uparrow} n_{i\downarrow} \quad (1),$$

where the first term accounts for the nearest-neighboring <ij> hopping, $c_{i\alpha}^\dagger$ ($c_{i\alpha}$) is the creation (annihilation) operator of a $J_{eff}$ = 1/2 electron with spin $\alpha$ on site $i$, $\sigma$ is the vector of Pauli matrices, and $U$ is the onsite electron-electron repulsion. The parameter $\theta$ in the SU(2) gauge field $e^{i\theta d_{ij} \cdot \sigma}$ controls the imaginary (spin-dependent) and real (spin-independent) hopping of a Ir-O-Ir bond by the ratio $\tan\theta$, whereas unit vector $d_{ij}$ stands for the spin rotation axis between the two Ir sites and hence depends on oxygen displacement of the bond [46]. However, a large $\theta$ does not guarantee Berry phases. For instance, c-axis octahedral rotations yield $d_{ij} = (-1)^i \hat{z}$, which interestingly allows the SU(2) gauge fields of all bonds to be gauged away simultaneously by a staggered frame transformation (Fig. 1a) [35, 43, 46]. The gauge-invariant flux $\phi$ of any closed path of hopping is thus zero the same as that when $\theta = 0$ [5]. This property refers to *the hidden SU(2) symmetry* of this "twisted" Hubbard Hamiltonian, which is preserved even with further-neighbor hopping [43].

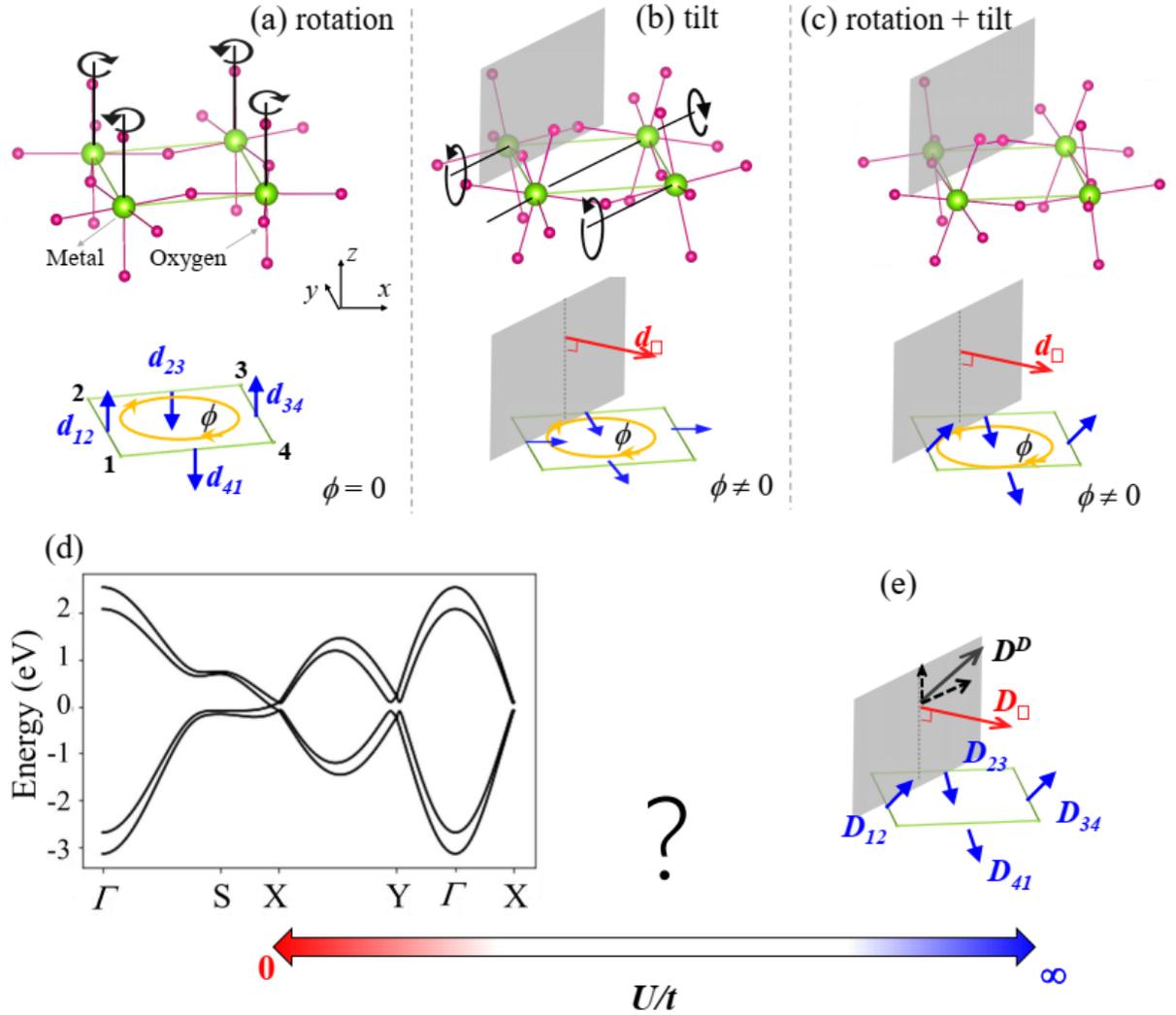

**FIG. 1. Model analysis and structure characterization.** A perovskite square lattice with (a) octahedral rotations (*p4g*), (b) octahedral tilts (*pg*), or (c) octahedral rotations and tilts (*pg*). Bottom panels show the modulation of $\boldsymbol{d}_{ij}$-vectors (blue arrows). Octahedral rotation/tilts on neighboring sites are opposite. The finite $\boldsymbol{d}_\square$ (red arrow) in (b) and (c) is perpendicular to the mirror plane (grey) of the *b*-glide symmetry. (d) Representative $U = 0$ band structure for a quasi-2D structure with two layers of square lattice (c) [47]. (e) Schematic diagram of DM interactions on the square lattice (c) at the strong coupling limit $U = \infty$. The blank space with a question mark highlights the extensive and largely unexplored intermediate region among the two limits.

Even more interesting situations thus arise if the local bond-inversion symmetry-breaking also breaks the hidden SU(2) symmetry [48]. As shown in Fig. 1b, we consider here octahedral tilt around a diagonal axis of the square plaquette, which causes the same $\sqrt{2} \times \sqrt{2}$ supercell expansion as the rotation while removing the four-fold rotation symmetry [49]. Thus, although all the bonds still have the same $\theta$, the $\boldsymbol{d}_{ij}$-vectors on the *x*- and *y*-bonds become noncollinear and rotated by 90º from each other due to the remaining glide plane. When both rotation and tilt are present, the $\boldsymbol{d}_{ij}$ vectors become non-coplanar with both in-plane and out-of-plane components (Fig. 1c), and the gauge field has both SU(2)-symmetry-breaking and preserving components. Nonetheless, only the SU(2)-symmetry-breaking component contributes to the finite $\phi$ around a plaquette, because $\phi$, to the first order of $\theta$, scales with the gauge-invariant vector $\boldsymbol{d}_\square \equiv \sum_\square \boldsymbol{d}_{ij}$ (where $\square$ stands for the consecutive sum over a plaquette) which is normal to the glide plane (Fig. 1c). Such a 2D lattice corresponds to the *pg* wallpaper group.

In the absence of correlation ($U = 0$), tight-binding Hamiltonian calculations indeed predicted such a 2D system to be a nonsymmorphic Dirac semimetal with Dirac points near the X and Y points of the unfolded Brillouin zone [45]. Figure 1d shows the $J_{\text{eff}} = 1/2$ band dispersion when stacking such square lattices into a quasi-2D structure, which has been proposed to host a quantum anomalous Hall phase upon breaking time-reversal-symmetry by a magnetic order [45]. Magnetism of the $J_{\text{eff}} = 1/2$ electrons is however often understood in the large-$U/t$ limit by mapping Eq.1 onto a Heisenberg-like model where the real hopping leads to the antiferromagnetic (AFM) Heisenberg superexchange interaction and the imaginary hopping is converted into the anisotropic superexchange interaction described by the Moriya vector $\boldsymbol{D}_{ij} = 4t^2/U \cdot \sin 2\theta \cdot \boldsymbol{d}_{ij}$ (Fig. 1e). Breaking the hidden SU(2) symmetry then implies that $\boldsymbol{D}_{ij}$ [50-52] has a finite gauge-invariant component characterized by $\boldsymbol{D}_\square \equiv \sum_\square \boldsymbol{D}_{ij} = 4t^2/U \cdot \sin 2\theta \cdot \boldsymbol{d}_\square$, and a gauge-dependent component described by the Dzyaloshinskii vector [46, 53] $\boldsymbol{D}^D \equiv \sum_j \boldsymbol{D}_{ij}$ in the global frame. While $\boldsymbol{D}^D$ is usually referred as the Dzyaloshinskii-Moriya interaction and can be gauged away, nonzero $\boldsymbol{D}_\square$ would induce Ising anisotropy along the $\boldsymbol{d}_\square$ axis [50] (Fig. 1e). It is clear that the SU(2) gauge field of the Hubbard model could lead to nontrivial and drastically different consequences in the weak and strong coupling limits, which are necessarily connected by a Dirac semimetal-to-Mott insulator transition in the present case. It is crucial to determine whether the topological and magnetic effects of the SU(2) gauge field could coexist in the intermediate coupling regime or in the vicinity of the transition and to what extent they could still be described by the pictures of the respective limits.

**Implementing $\phi$ via artificial synthesis**

The IrO$_6$ octahedral layer is an ideal building block for simulating such a Hubbard Hamiltonian not only because a large $\theta$ can be afford by the SOC but also due to the intermediate-$U/t$ value of 5$d$ electrons [30, 33, 54, 55]. The final flavor to realize such a toy-model material is the control of the rotation and tilt, which we achieved by growing [(SrIrO$_3$)$_1$/(CaTiO$_3$)$_1$] superlattices (SLs) (Fig. 2a) with 30 repeats on (001)-oriented SrTiO$_3$ single-crystal substrates by pulsed laser deposition (Materials and Methods). This design implements significant octahedral tilt, which is inspired by the idea of reducing the effective tolerance factor of the reported rotation-only [(SrIrO$_3$)$_1$/(SrTiO$_3$)$_1$] SL [56, 57] by replacing the confining SrTiO$_3$ monolayers with CaTiO$_3$. We defined the reciprocal lattice based on an $a \times a \times 2c$ supercell, where $a$=3.905 Å and $c$=3.892 Å are the pseudo-cubic in-plane and out-of-plane lattice parameters, respectively, extracted from x-ray diffraction (XRD). Figure 2b shows the primary (0 0 *even*) film peaks, which almost overlap with the substrate peaks, as well as the (0 0 3) satellite reflection that confirms the layered structure (Fig. 2a). A fully strained state is also confirmed by reciprocal space mapping, shown in Supplemental Fig. S1 [47]. The octahedral pattern was verified by measuring a set of half-order peaks. For instance, showing in Fig. 2c is a pronounced (0.5 0.5 3) reflection characteristics of significant octahedral tilt [58], confirming the successful implementation of the designed SU(2)-symmetry-breaking. The (0.5 1.5 3) peak (Fig. 2d) representing octahedral rotation is also observed, shown in Supplemental Fig. S3 [47].

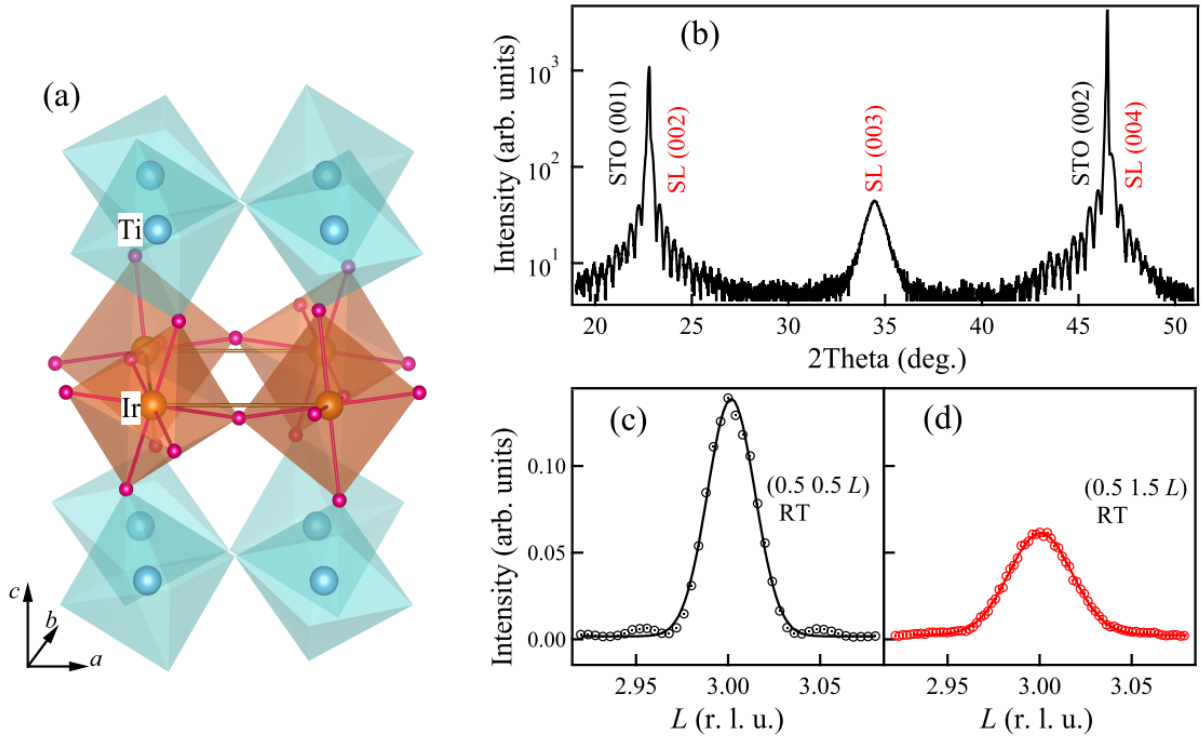

**FIG. 2. Structure characterization.** (a) Crystal structure of the [(SrIrO$_3$)$_1$/(CaTiO$_3$)$_1$] SL. (b) X-ray $\theta$-$2\theta$ scan pattern of the SL. (c)&(d) Synchrotron XRD measured around the (0.5 0.5 3) and (0.5 1.5 3) peaks.

**Mott state of intermediate coupling regime**

We first established the electronic ground state of the SL as a Mott insulator by measuring the longitudinal resistivity that increases with lowering temperature (Fig. 3c), and confirmed the square-lattice AFM order by probing the (0.5 0.5 *integer*) AFM Bragg peak with resonant x-ray magnetic scattering (Fig. 3a). The AFM order in one $IrO_6$ octahedral layer is thus related to another simply by vertical translation. The integrated peak intensity recorded as a function of temperature reveals $T_N \sim 170$ K (Fig. 3b). These results conclude that the correlation strength of $[(SrIrO_3)_1/(CaTiO_3)_1]$ is sufficiently large to turn the nonsymmorphic Dirac semimetallic state in the weak coupling limit into an AFM Mott insulating state.

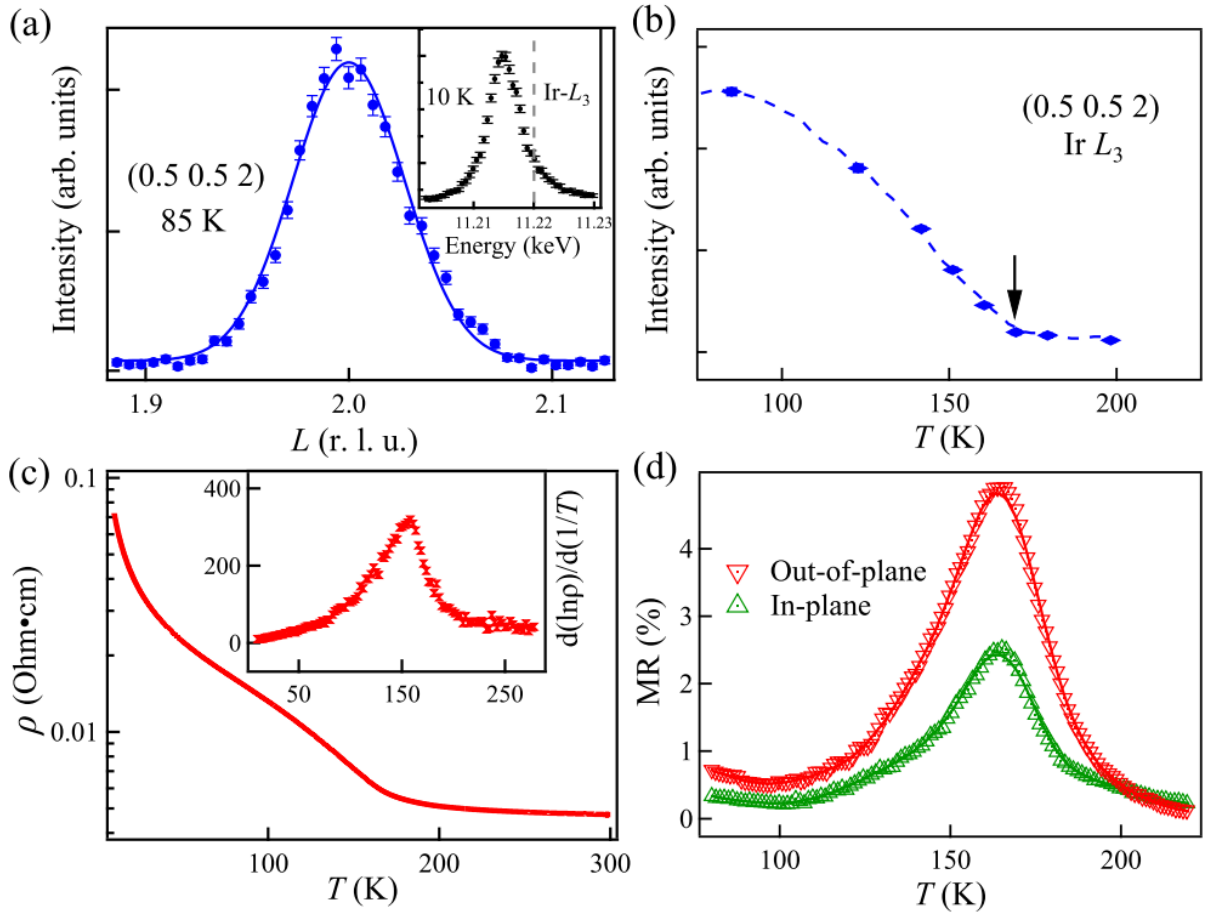

**FIG. 3. Magnetic structure and Magnetoresistance effect.** (a) *L*-scan across the (0.5 0.5 2) magnetic peak. Error bar represents statistic error. Inset shows the energy profile at 10 K, where a dashed line indicates the Ir-$L_3$ absorption edge. (b) Integrated intensity of magnetic peak versus temperature. $T_N$ is denoted by a black arrow. (c) Temperature dependence of the in-plane resistivity. Inset shows the relation between $d(ln\rho)/d(1/T)$ and temperature. (d) Temperature dependent MR measured under an out-of-plane (red) or in-plane (green) 8 T magnetic field.

Meanwhile, there are also clear signatures that the $U/t$ value of [(SrIrO$_3$)$_1$/(CaTiO$_3$)$_1$] falls into the intermediate regime, such as a resistivity kink right below $T_N$ (Fig. 3c) which implies significant charge fluctuations that are suppressed upon long-rang AFM ordering [33]. Suppressing these charge fluctuations by a magnetic field results in a positive anomalous magnetoresistance (MR) with a critical enhancement in the paramagnetic insulating phase (Fig. 3d). This magneto-charge transport response above $T_N$ is due to the fact that the charge fluctuations are tied with the longitudinal AFM fluctuations and the external uniform field suppresses them by acting as an effective staggered field and inducing finite staggered magnetization. This staggered field effect is in virtue of the gauge-dependent component of the imaginary hopping in Eq.1 [33]. Specifically, the response to the in-plane field is attributed to the out-of-plane component of the $\bm{d}_{ij}$ vectors [59] that originates from the octahedral rotation and can be gauged away by a staggered rotation transformation around the c-axis. The response to the out-of-plane field, on the other hand, demonstrates that the in-plane component of the $\bm{d}_{ij}$ vectors (induced by the tilt) also contributes to the gauge-dependent imaginary hopping as shown in Figs.1(a&b). All in all, the significant field-tunable spin-charge fluctuations are unexpected for a strong Mott insulator but characteristics of the intermediate coupling regime where the correlated charge gap allows excitations into the electron-hole continuum.

**Hall signature of topology**

In addition to the anomalous MR, the charge fluctuations also manifest in the emergence of a spontaneous Hall effect upon broken time-reversal symmetry in $[(SrIrO_3)_1/(CaTiO_3)_1]$. One can see a nonlinear Hall signal with the out-of-plane field developing around $T_N$, and it evolves into a hysteresis loop with a significant remnant Hall resistance upon cooling (Fig. 4a). While such an AHE is a signature of significant Berry curvatures which are consistent with the Dirac points predicted in the non-interacting band structure (Fig. 1d), its emergence within the Mott insulating state clearly shows the survival of this topological effect in the Hubbard bands. This gives rise to unique temperature dependent behaviors as seen in Fig. 4b that both the extracted saturated and remnant values of the AHE conductivity start to decrease with cooling after a sharp increase below $T_N$.

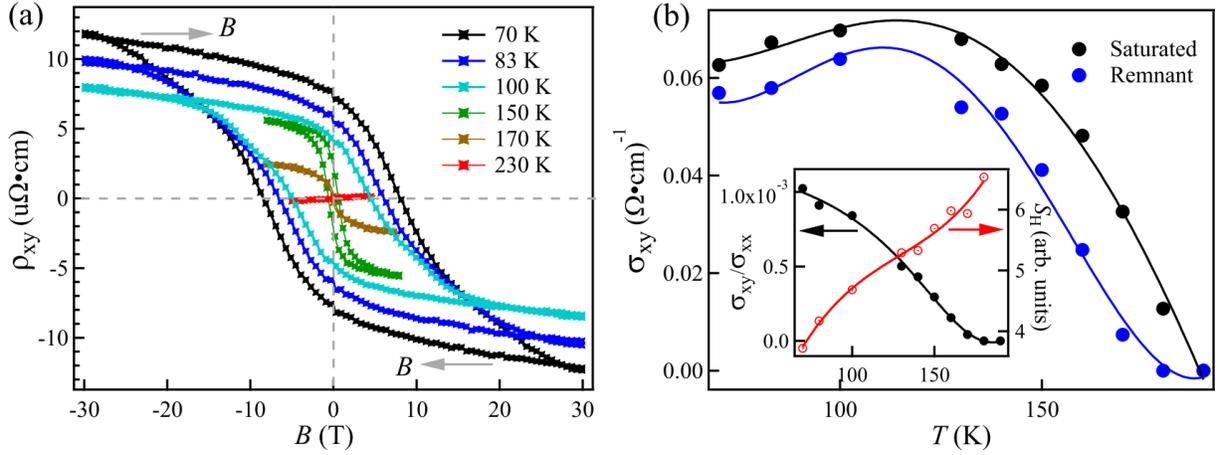

**FIG. 4. Hall signatures of the topology.** (a) Magnetic field dependence of Hall resistivity at different temperatures. The field sweeping directions are shown as grey arrows. (b) Temperature dependences of the saturated and remnant Hall conductivity. Inset shows the temperature dependences of $\sigma_{xy}/\sigma_{xx}$ and $S_H$ [47].

This behavior is a result of the self-competition of the electron-hole pairing since the AHE conductivity relies on thermally excited charge carriers in the electron-hole continuum as well as the time-reversal symmetry-breaking field from the AFM order. However, the AFM order parameter itself necessarily increases at the expense of the longitudinal spin fluctuations by reinforcing the electron-hole binding. Such competition is captured by the anomalous Hall angle $\sigma_{xy}/\sigma_{xx}$ and the anomalous Hall conductivity coefficient $S_H = \sigma_{xy}/M_s$ [60], where $M_s$ is the staggered magnetization (Inset of Fig.4b). The former indicates that the current is deflected more and more upon cooling (i.e., the magnetic order becomes stronger and stronger), while the latter suggests that the charge carriers contributing to the transverse channel drop quickly below $T_N$. These observations highlight the topology-correlation interplay in the intermediate coupling regime of the Hubbard Hamiltonian. In other words, the observed AHE is completely driven by the same group electrons (i.e., the $J_{eff} = 1/2$ electrons) that form the AFM Mott insulating state, which is in sharp contrast to the AHE reported in iridate/manganite and iridate/ruthenate heterostructures, where the strong spin-orbit coupled 5$d$ electrons and the itinerant 3$d$ and 4$d$ electrons play separate roles in the physical properties [61-64].

**Emerging Ising anisotropy**

The spontaneous Hall conductivity indicates that the mirror of the glide plane symmetry must be broken by the magnetic order, which is actually consistent with the expectation in the strong coupling limit that the AFM spin axis is along the $D_□$ vector perpendicular to the glide plane (Fig. 1e & inset of Fig. 5a) [50-52]. While $D_□$ originates from the gauge-invariant component of the imaginary hopping [46, 53], the gauge-dependent component $D^D$ necessarily introduces spin canting of the AFM order [50]. However, the $D^D$ vector alone does not specify a canting direction because of the rotational invariant nature of the DM interaction. The canting direction and angle are instead determined by $\boldsymbol{D}_\square \times \boldsymbol{D}$ [50]. Under the implemented rotation and tilt, $\boldsymbol{D}_\square$ and $\boldsymbol{D}^D$ are normal to each other (Fig.1e), which should give rise to both in-plane and out-of-plane canted moments. Our magnetization measurement (Fig.5a) indeed shows both in-plane and out-of-plane remnant magnetizations below $T_N$.

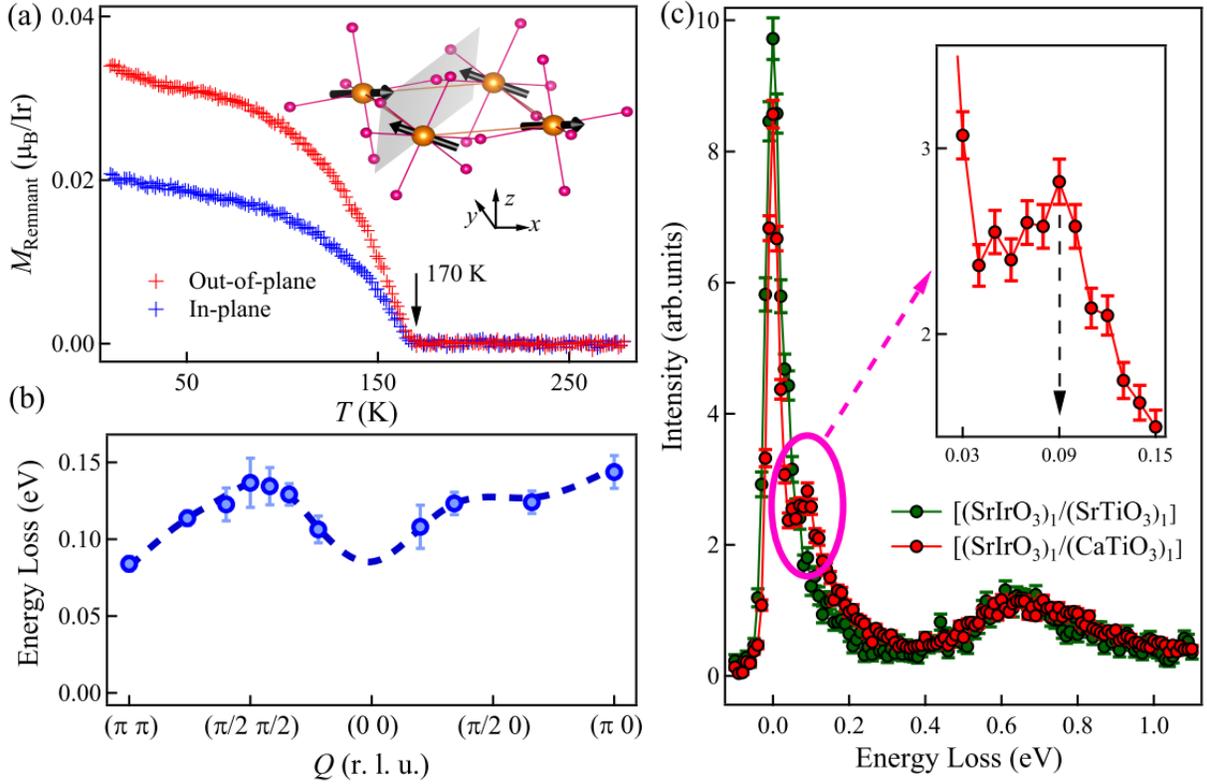

**FIG. 5. Magnetic structure and low-energy magnetic excitations.** (a) Temperature dependent out-of-plane (red) and in-plane (blue) remnant magnetization ($M_{Remnant}$). Inset shows the ground-state magnetic pattern of the SL. The $J_{eff} = 1/2$ magnetic moments are represented as black arrows. (b) Extracted magnon peak dispersion across the Brillion zone. The errors are statistical from the fitting with dashed line as a guide to the eye. (c) RIXS spectra at $\boldsymbol{Q} = (\pi, \pi)$ of the SL

(red) and [(SrIrO$_3$)$_1$/(SrTiO$_3$)$_1$] (green) measured at 35 K. The magnon excitation feature is indicated as an arrow. Inset shows the expanded view of the magnon feature.

While correctly capturing the symmetry of the magnetic structure, we found that the superexchange approach is however insufficient to account for the strength of the Ising anisotropy, which we evaluated through resonant inelastic x-ray scattering measurements at the Ir $L_3$-edge. The technique has been widely used to measure magnon excitations in iridates [65-69], including thin films [69] and artificial SLs [70, 71]. To map the dispersion of magnetic excitations, RIXS spectra were collected at multiple $Q$ points. Indeed, we found the overall magnon dispersion in [(SrIrO$_3$)$_1$/(CaTiO$_3$)$_1$] is similar to other single-layered iridates [47] and characteristic of a square-lattice Heisenberg antiferromagnet except near the AFM wave vector ($\pi$, $\pi$) [70, 72, 73]. Figure 5c shows the energy-loss spectra at ($\pi$, $\pi$), displaying a sharp elastic peak at 0 eV and a broad bump centering at ~0.65 eV due to the overlap of spin-orbit excitation and particle-hole continuum [70, 72, 73], which is similar to other iridates. From 0.05 to 0.14 eV, a prominent magnon excitation is well separated from the elastic peak. For comparison, the spectrum of [(SrIrO$_3$)$_1$/(SrTiO$_3$)$_1$] at the same $Q$ point is overlaid to highlight the difference due to the breaking of the hidden SU(2) symmetry by the implemented tilt distortion. As showing in Fig. 5b, the extracted magnon peak at ($\pi$, $\pi$) along with the dispersion gives a giant spin gap $\Delta_m$~85 $\pm$ 5 meV. Given a Heisenberg $J$~50 meV in most iridates [70-73], this large value of $\Delta_m$ would yield $D_\square$~160 meV in the linear spin wave theory, which effectively requires $U/t$ ~ 1 and invalidates the superexchange approach that assumes a large-$U/t$ limit[47]. It again points to the fact that the system is in the intermediate coupling regime, and the giant spin gap is a nontrivial and direct result of the spatial fluctuations of the electron-hole pairs through the SU(2)- symmetry-breaking hopping. A theoretical estimation of the spin gap should be done directly with the Hubbard model and indeed suggests $U/t$~2.5 for such a large magnon gap [74].

To conclude, we have experimentally realized a quasi-2D anomalous Hall Mott insulator, which is designed to incorporate correlated and topological effects in the square-lattice Hubbard Hamiltonian with a frozen SU(2) gauge field. The results show that the gauge-dependent and gauge-invariant components of the SU(2) gauge field lead to distinct consequences that coexist in the intermediate correlation regime. In particular, the gauge-invariant component gives rise to an emergent spontaneous Hall effect, which is a signature of Berry curvatures in the electronic structure of a Mott state derived from a Dirac semimetallic phase. The electronic correlation of the Mott state in turn renders the spontaneous Hall effect a nonmonotonic temperature dependent behavior as the AFM order is necessarily stabilized at the expense of the spatial charge fluctuations that the Hall effect relies on. On the other hand, the spin axis of the AFM order is determined by the gauge-invariant component which creates a surprisingly strong Ising anisotropy beyond the conventional superexchange approach based on the Moriya vector. The intertwining of phenomena that are usually captured in drastically different pictures of the electronic state highlights the rich and complex interplay between correlation and topology in the intermediate coupling regime.

## Materials and Methods

**Sample preparation.** During the sample growth, the substrate temperature and laser fluence (KrF excimer laser beam, λ = 248 nm) were optimized to be 600°C and 2 J/cm$^2$, respectively. The growth pressure was maintained to be 100 mbar with a constant oxygen flow. A reflection high energy electron diffraction (RHEED) unit was adopted to accurately control the stacking sequence and SL thickness (30 SL unit cells).

**Experimental methods.** The crystalline quality and lattice structure of the obtained SL were characterized on a Panalytical X'Pert MRD diffractometer. Electric transport measurements were carried out on a physical properties measurement system (Quantum design). High-field (30 T) Hall measurement was performed on the Steady High Magnetic Field Facilities, High Magnetic Field Laboratory, Chinese Academy of Sciences. Measurements of magnetization were performed on a vibrating sample magnetometer (Quantum design). Synchrotron x-ray diffraction experiments were performed at 33BM beamline in the Advanced Photon Source (APS) of Argonne National Laboratory. X-ray absorption measurement was performed at 4IDD beamline, APS. Magnetic resonant X-ray scattering experiment was carried out at 6IDB beamline, APS. A polarization analyzer was utilized to increase the signal-to-noise ratio. The reciprocal lattice notation is defined based on a $a \times a \times 2c$ superlattice cell, where $a$ and $c$ are the pseudo-cubic in-plane and out-of-plane lattice parameters, respectively. The interlayer coherence length can be extracted to be ~$64c$ from Fig. 2A in the main text, indicating that the quasi-2D ordering spans over the entire SL. RIXS experiments were conducted at 27IDB beamline, APS, with a grazing incidence geometry. The reciprocal space resolution is ~ 0.12 Å$^{-1}$ with a mask used.

**Acknowledgement:** The authors acknowledge experimental assistance from H.D. Zhou and M. Koehler. The authors also appreciate helpful discussions with C. D. Batista, E. Dagotto, S. S. Zhang, Z. T. Wang and D. Gong. **Funding:** J. L. acknowledges support from the National Science Foundation under Grant No. DMR-1848269 and the Office of Naval Research (Grant No. N00014-20-1-2809). H.S. acknowledges Inamori Research Grants from Inamori Foundation and support from JSPS KAKENHI Grants No. JP19K14650 and JP22K03508. Lin Hao acknowledges support from the National Natural Science Foundation of China (Grant No. 12104460) and the High Magnetic Field Laboratory of Anhui Province (Grant No. AHHM-FX-2021-03). Lukas Horak acknowledges the support by the ERDF (project CZ.02.1.01/0.0/0.0/15_003/0000485) and the Grant Agency of the Czech Republic grant (14-37427 G). Work at Brookhaven National Laboratory was supported by the U.S. Department of Energy, Office of Science, Office of Basic Energy Sciences, under Contract No. DE-SC0012704. Use of the Advanced Photon Source, an Office of Science User Facility operated for the US

DOE, OS by Argonne National Laboratory, was supported by the U. S. DOE under contract no. DE-AC02-06CH11357. Part of characterization in this research was conducted at the Center for Nanophase Materials Sciences, which is a DOE Office of Science User Facility. A portion of this work was performed on the Steady High Magnetic Field Facilities, High Magnetic Field Laboratory, Chinese Academy of Sciences, and supported by the High Magnetic Field Laboratory of Anhui Province.

**Author contributions:** J.L. and L.H. conceived and directed the study. L.H. and J.Y. undertook sample growth and characterization. L.H., J.Y., Lukas Horak and J.K. performed structure characterization. L.H. and Z.S.W. performed high-field transport measurement. L.H., J.Y., D.M., J.W.K and P.J.R. performed magnetic scattering measurements. L.H., J.Y., G.F., Y.S.C. and D.H. conducted XAS measurements. D.M. and M.P.M.D. performed RIXS measurements. H.S. performed the theoretical simulation. L.H. and J.L. analyzed data. L.H. and J.L. wrote the manuscript.

**Competing interests:** The authors declare that they have no competing interests.

**Data and materials availability:** All data needed to evaluate the conclusions in the paper are present in the paper and/or the Supplementary Materials. Additional data related to this paper may be requested from the authors.